\documentclass[12pt,legalpaper,notitlepage]{article}
\usepackage{epsfig}
\date{}

\begin{document}
\vspace*{4.5cm}
\noindent{\Large\bf THE ISOTOPE EFFECT}\\[0.5cm]
{\Large\bf IN SUPERCONDUCTORS}\\[1cm]
\hspace*{1in}A.~Bill,$^1$ V.Z.~Kresin,$^1$ and S.A.~Wolf$^2$\\[0.5cm]
\hspace*{1in}$^1$ Lawrence Berkeley Laboratory, University of California,\\
\hspace*{1.1in}Berkeley, CA 94720, USA\\
\hspace*{1in}$^2$ Naval Research Laboratory, Washington D.C.~20375-5343 \\[.3cm]

\begin{abstract}
We review some aspects of the isotope effect (IE) in superconductors.
Our focus is on the influence of factors not related to the
pairing mechanism. After summarizing the main results obtained for
conventional superconductors, we review the effect of magnetic
impurities, the proximity effect and non-adiabaticity on the
value of the isotope coefficient (IC). We discuss the isotope effect
of $T_c$ and of the penetration depth $\delta$. The theory is
applied to conventional and high-$T_c$ superconductors. Experimental
results obtained for YBa$_2$Cu$_3$O$_{7-\delta}$ related materials
(Zn and Pr-substituted as well as oxygen-depleted systems)
and for La$_{2-x}$Sr$_x$CuO$_4$ are discussed.
\end{abstract}

\section{INTRODUCTION}\label{sec:intr}

Historically, the isotope effect (IE) played a major role in
unravelling the questions related to the origin of the effective
attractive interaction
between charge-carriers which leads to the superconducting
state. The theoretical considerations of Fr\"ohlich\cite{frohlich}
and the experimental discovery of the IE of $T_c$ in
mercury\cite{maxwell} pointed towards the contribution of lattice
dynamics to the instability of the normal state. Since then, the
IE has often been considered as a measure of the
contribution of phonons to the pairing mechanism. Furthermore, it
is generally assumed that only those thermodynamical quantities that
depend explicitely on the phonon frequencies (as $T_c$ or the order
parameter $\Delta$ in the BCS model) display an IE.

One can show\cite{KW1,KBWO,BKW1,BKW2,KW2}, however, that such understanding of
the isotope effect is incomplete and leads, therefore,
to confusion; this is particularly true for the high-temperature
oxides. Indeed, several factors not
related to the pairing mechanism can alter the value of the isotope
coefficient (IC).
Moreover, these factors are not necessarily
related to lattice dynamics. Here we focus our attention on three
such factors: magnetic impurities, the proximity effect and
non-adiabatic charge-transfer as it occurs in high-$T_c$
superconductors.

In addition, we also show \cite{BKW2} that fundamental
quantities such as the penetration depth $\delta$ of a magnetic
field also display an IE because of the three factors mentioned
above. Note that this effect occurs despite the fact that $\delta$
does not explicitely depend on quantities related to lattice
dynamics.

It results from our considerations that the value of the IC (even
its absence) does not allow any {\it a priori} conclusion about the
pairing mechanism. Nevertheless, it remains an interesting effect
that enables one to
determine the presence of magnetic impurities, proximity effect
or non-adiabaticity in the system. The calculation of the IC and
its comparison with experimental results was also used in previous
works to determine the value of the Coulomb repulsion $\mu^{\star}$
or the relative weight of different electron-phonon coupling
strengths in superconductors (see below). It can thus
be used as a tool for the characterization of superconductors.

This review is mainly based on our
papers\cite{KW1,KBWO,BKW1,BKW2,KW2}. We apply the theory to analyse
the oxygen isotope effect of
$T_c$ in Zn and Pr-doped (YBCZnO and YPrBCO) as well as
oxygen-depleted YBa$_2$Cu$_3$O$_{7-\delta}$ (YBCO).
We also review calculations of the penetration depth isotope effect.
The latter superconducting property is a good example of
a quantity that does not directly depend on phonon frequencies and
yet can display a substantial isotopic shift \cite{BKW2}. In
this context, we also discuss recent experimental results obtained
for La$_{2-x}$Sr$_x$CuO$_4$ (LSCO).

The structure of the paper is as follows. In section
\ref{sec:HIST} we present some of the early results related to the
isotope effect and show that even for conventional superconductors
the relation between the isotope effect and the pairing mechanism
is not simple. We also discuss the applicability of these early
results to the description of high-temperature superconductors.

The remaining sections are devoted to the description of new,
unconventional isotope effects. In section
\ref{sec:MI} we study the influence of magnetic impurities on the
value of the isotope effect. We show that adding magnetic impurities
to a superconductor can
enhance the isotope coefficient $\alpha$ of $T_c$ and induce a
temperature-dependent isotope effect of $\delta$.
We show that both isotope effects are universal functions of $T_c$.
That is, $\alpha(T_c)$ and $\beta(T_c)$ are independent of any
adjustable parameter.
We discuss Zn-doped YBa$_2$Cu$_3$O$_{7-\delta}$ in this context.

Section \ref{sec:prox} is concerned with the influence of
a normal layer on the isotope effect of a superconductor. We
show that due to the proximity effect, the isotope coefficient
of $T_c$ is linear in the ratio of the normal to the superconducting
film thicknesses $\rho$. Furthermore, the proximity effect induces a
temperature and $\rho$-dependent isotope coefficient of the
penetration depth.

Section \ref{sec:NA} reviews the concept of the
non-adiabatic isotope effect introduced in Ref.~\cite{KW1} and
further discussed in Refs.~\cite{KBWO,BKW1,BKW2,KW2}. We show that
in systems as the high-temperature superconductors where
charge-transfer processes
between reservoir and CuO$_2$-planes occur via ions that
display a non-adiabatic behaviour, the charge-carrier density in
the planes depend on the ionic mass. This leads to the unconventional
non-adiabatic contribution to the IE of $T_c$. It is also interesting
that non-adiabaticity induces an isotopic shift of the penetration
depth $\delta$.

In the last section (Sec.~\ref{sec:HTSC}) we apply our theory to
the oxygen
isotope effect (OIE) in high-temperature superconductors of the
YBCO-family. Focus is set on the oxygen isotope effect because most
experimental studies have been performed on the oxygen ion which is
the lightest in the CuO$_2$ plane. The situation is less clear in
the case of copper isotopic substitution where an effect
has also been observed\cite{franckCu,zhaoCu}. This latter case will
be discussed in more detail elswhere.

We conclude the review in section \ref{sec:concl}.

\section{THE ISOTOPE EFFECT: DISCOVERY,\\CONVENTIONAL VIEW}
\label{sec:HIST}

The isotope effect of superconducting critical temperature $T_c$ is
best described in terms of the isotope
coefficient (IC) $\alpha$ defined by the relation
$T_c \sim M^{-\alpha}$, where $M$ is the ionic mass. Under the
assumption that the shift $\Delta T_c$ induced by isotopic
substitution ($M\rightarrow M^{\star}$) is small compared to $T_c$,
one can write
\begin{eqnarray}\label{ICTc}
\alpha = - \frac{M}{\Delta M} \frac{\Delta T_c}{T_c} \qquad ,
\end{eqnarray}
where $\Delta M = M-M^{\star}$ is the difference between the two
isotopic mass.
If the superconductor is composed of different elements, one defines
a {\it partial} isotope coefficient $\alpha_r$ as in
Eq.~(\ref{ICTc}), but where $M$ is replaced by $M_r$, the mass
of element $r$ that is substituted for its isotope. In addition,
one defines the total isotope coefficient by
\begin{eqnarray}\label{ICTcpartial}
\alpha_{tot} = \sum_r \alpha_r =
- \sum_r \frac{M_r}{\Delta M_r} \frac{\Delta T_c}{T_c} \qquad .
\end{eqnarray}

Besides $T_c$ there are other quantities that display an isotope
effect. In the next sections we focus on the isotope shift of the
penetration depth $\delta$. In analogy to the IC of $T_c$ (denoted
$\alpha$) we define the isotope coefficient $\beta$ of the
penetration depth by the relation
\begin{eqnarray}\label{ICdelta}
\beta =- \frac{M}{\Delta M} \frac{\Delta \delta}{\delta} \qquad .
\end{eqnarray}
One should note that there is a conventional BCS-type isotope effect
of the penetration depth related to its temperature dependence
$\delta^{-2} \sim (1-T^4/T_c^4)$. Because
$T_c$ displays an isotope effect, the penetration depth is also
shifted upon isotopic substitution. The effect becomes strong as one
approaches $T_c$. The present article is not concerned with this
trivial effect.

Let us first discuss what values of the isotope coefficients
$\alpha$ and $\beta$ can be expected for different types of
superconductors. Table 1 shows characteristic
values of the isotope coefficient for different types of
superconductors.
Very different values of the IC have been observed
(in the range from $-2$
to $+1$). One notes that some systems have a negligible coefficient,
some display even an ``inverse isotope coefficient'' ($\alpha < 0$)
and some take values greater than $0.5$, the value predicted
by Fr\"ohlich and the BCS model (for a monoatomic system). The
purpose of the next paragraphs is to describe shortly different
theoretical models allowing one to understand
the coefficients observed (see Table 1 and
Refs.~\cite{gladstone,lynton}). We begin with the description of
conventional superconductors and discuss the relevance of these
models for high-$T_c$ materials. In the following sections we then
introduce new considerations about the IE allowing one to give a
consistent picture of the IE in high-temperature superconductors.
\newpage
\noindent{\bf Table 1.} Experimental values of the isotope coefficient of $T_c$ (see also Refs.~\cite{lynton,meservey}). The letters in the last column correspond to Ref.~\cite{maxwell}.\\
\nopagebreak[4]
\begin{center}
{\small \begin{tabular}{llc}\hline
Superconductor	& $\alpha$ & Reference (see \cite{maxwell})\\\hline
Hg	&	$0.5 \pm 0.03$	&	a \\
Tl	&	$0.5 \pm 0.1$	&	a, and \cite{lynton}\\
Cd	&	$0.5 \pm 0.1$	&	b\\
Mo	&	$0.33 \pm 0.05$	&	c\\
Os	&	$0.21 \pm 0.05$	&	d,e\\
Ru	&	$0.0$		&	e\\
Zr	&	$0.0$		&	e\\
PdH(D)	&	$-0.25$		&	\cite{stritzker,ganguly}\\
U	&	$-2$		&	\cite{fowler}\\
La$_{1.85}$Sr$_{0.15}$CuO$_4$& $0.07$&	\cite{crawford}\\
La$_{1.89}$Sr$_{0.11}$CuO$_4$& $0.75$&	\cite{crawford}\\
($^{16}$O $-$ $^{18}$O subst.)&&\\
K$_3$C$_{60}$	& $0.37$ or $1.4$ &	\cite{ashcroft}\\
($^{12}$C $-$ $^{13}$C subst.)&&\\\hline
\end{tabular}}
\end{center}
\vspace*{0.5cm}

\subsection{Monoatomic Systems}

Let us begin with the case of a monoatomic BCS-type superconductor.
Substituting the atoms by an isotope affects the phonon dispersion
(for a monoatomic lattice $\Omega \sim M^{-1/2}$, where $\Omega$ is
a characteristic phonon frequency). Thus, any quantity that
depends on phonon frequencies is affected by isotopic substitution.
Assuming that the electron-electron pairing interaction is
mediated by phonons and neglecting the Coulomb repulsion between
electrons, the BCS theory predicts that $T_c\propto\Omega$ (see
below, Eq.~(\ref{TcBCS}) for $\mu^{\star} = 0$). Thus, for a
monoatomic system the isotope coefficient of $T_c$ given by
Eq.~(\ref{ICTc}) is $\alpha = 0.5$.

As seen in Table 1, the IC of most monoatomic non-transition metals
is approximately equal to $0.5$. Many other systems,
on the other hand, deviate from this value. In
particular, transition metals and alloys display values
that are smaller than, or equal to $0.5$. PdH displays an
inverse isotope effect $\alpha_{PdH} \simeq -0.25$ \cite{stritzker}
(there is also one report on a large inverse IC of Uranium
$\alpha_U \simeq -2$\cite{fowler}). Furthermore, high-$T_c$
materials can have values both smaller and larger than $0.5$,
depending on the doping.

\subsection{The Coulomb Interaction}\label{sssec:coulomb}

One of the first reasons advanced to explain the discrepancy
between theory ($\alpha = 0.5$) and experiment
was that the BCS calculation described above did not take properly
into account the Coulomb repulsion between charge-carriers.
It was shown in Refs.~\cite{bogoliubov,khalatnikov,swihart,morel}
that inclusion of Coulomb interactions leads to the introduction
of the pseudo-potential $\mu^{\star}$ in the BCS equation for $T_c$:
\begin{eqnarray}\label{TcBCS}
T_c = 1.13\Omega \exp{\left(- \frac{1}{\lambda - \mu^{\star}}\right)}
\qquad ,
\end{eqnarray}
where $\mu^{\star}$ is the pseudo-potential given by
\begin{eqnarray}\label{mustar}
\mu^{\star} =
\frac{\mu}{1 + \mu\ln\left(\frac{E_F}{\Omega}\right)} \qquad .
\end{eqnarray}
$\mu$ is the Coulomb potential and $E_F$ is the Fermi energy.
One notes that $\Omega$ is present in $\mu^{\star}$ and thus in the
exponent of Eq.~(\ref{TcBCS}). The value of $\alpha$ can thus be
substantially decreased from $0.5$. Indeed, from Eq.~(\ref{ICTc})
and (\ref{TcBCS}) one obtains
\begin{eqnarray}\label{ICTcBCS}
\alpha = \frac{1}{2}\left\{ 1 -
\left( \frac{\mu^{\star}}{\lambda - \mu^{\star}} \right)^2 \right\}
\qquad .
\end{eqnarray}
The isotope coefficient is shown as a function of $\lambda$ for
different values of $\mu^{\star}$ in Fig.~1. Naturally, only the
case $\lambda > \mu^{\star}$ is relevant, since superconductivity
is otherwise suppressed.

The effect of Coulomb interactions can be included in a similar way
in the strong-coupling Eliashberg theory. Using the
formula derived by McMillan \cite{mcmillan}:
\begin{eqnarray}\label{mcmillan}
T_c = \frac{\Omega}{1.2}
\exp{\left(- \frac{1.04(1+\lambda)}{\lambda - \tilde{\mu}}\right)},
\end{eqnarray}
with $\tilde{\mu} \equiv (1 + 0.62\lambda)\mu^{\star}$ and
$\mu^{\star}$ is given by Eq.~(\ref{mustar}), one obtains the
following result for the isotope coefficient:
\begin{eqnarray}\label{ICmcmillan}
\alpha = \frac{1}{2}
\left\{ 1 - \frac{1.04(1+\lambda)\tilde{\mu}\mu^{\star}}{\left[
\lambda - \tilde{\mu}\right]^2} \right\} \qquad .
\end{eqnarray}
A similar result has also been derived in Ref.~\cite{gladstone}.
Fig.~1 shows the dependence described by Eq.~(\ref{ICmcmillan})
and allows one to compare the weak and strong-coupling cases, that
is, Eqs.~(\ref{ICTcBCS}) and (\ref{ICmcmillan}).

An expression for $T_c$ that is valid over the whole range of
coupling strengths (from weak to very strong couplings) has been
derived by one of the authors in Ref.~\cite{kresinTc}:
\begin{eqnarray}\label{Tckresin}
T_c = \frac{0.25\Omega}{\sqrt{e^{2/\lambda_{eff}}-1}}
\end{eqnarray}
with
\begin{eqnarray}\label{lambdakresin}
\lambda_{eff} = \frac{\lambda - \mu^{\star}}{1+2\mu^{\star}
+ \lambda\mu^{\star}t(\lambda)} \qquad,
\end{eqnarray}
and $t(\lambda) \simeq 1.5 exp(-0.28\lambda)$\cite{kresinTc}. As
usual, the parameter $\mu^{\star}$ is given by Eq.~(\ref{mustar}).
The isotope coefficient resulting from Eq.~(\ref{ICTc}),
(\ref{Tckresin}) and (\ref{lambdakresin}) reads
\begin{eqnarray}\label{ICkresin}
\alpha = \frac{1}{2}\left\{
1 - \frac{{\mu^{\star}}^2}{\lambda_{eff}\left(1-e^{-2/\lambda_{eff}}\right)}
\left[ \frac{1}{\lambda - \mu^{\star}} + 
\frac{2 + \lambda t(\lambda)}{3 + \lambda t(\lambda)} \right]
\right\} \qquad .
\end{eqnarray}
This isotope coefficient is also displayed in Fig.~1 for two
values of the parameter $\mu^{\star}$.
\unitlength1cm
\begin{figure}[htb]
\begin{center}
\input{IC_BMK.ps}
\end{center}
{\small {\bf Figure 1.} Isotope coefficient $\alpha$ as a function of $T_c/\Omega$ (i.e.~$\lambda$) for different values of $\mu^{\star}$. Solid lines from Eq.~(\ref{ICTcBCS}), dashed lines from Eq.~(\ref{ICmcmillan}) and dash-dotted lines from Eq.~(\ref{ICkresin}). For each type of line, the upper curve corresponds to $\mu^{\star} = 0.1$ and the lower to $\mu^{\star} = 0.2$.}
\end{figure}
Note that for all models, because of the presence of $\mu^{\star}$
the isotope coefficient of $T_c$ is lowered
with respect to the BCS value for $\alpha = 0.5$.
One recovers the BCS result asymptotically at large $\lambda$ (or
$T_c/\Omega$). Only Eq.~(\ref{ICkresin}), however, is valid in the
strong coupling limit ($\lambda>1$).

An interesting general feature of the results presented
above is that the strongest
deviation from the BCS monoatomic value $0.5$ occurs when
$\mu^{\star}$ (or $\tilde{\mu}$) is of the order of $\lambda$
(see Eqs.~(\ref{ICTcBCS}), (\ref{ICmcmillan}) or (\ref{ICkresin})).
Thus, excluding
anharmonic or band-structure effects, {\it a small} isotope
coefficient is correlated to a {\it low} $T_c$. This is indeed
observed in most conventional superconductors. In certain cases one
even obtains a {\it negative} (also called {\it inverse}) isotope
coefficient ($\alpha < 0$). For realistic values of the parameters,
$T_c$ should not exceed $T_c\sim 1$K for this inverse effect to
be present. Several systems display this inverse isotope effect,
among them PdH (where H is replaced by its isotope D)
\cite{stritzker,ganguly} and Uranium (where $^{235}$U is replaced by
$^{238}$U) \cite{fowler}. We discuss this case
below (sec.~\ref{sssec:band}).

The situation encountered in high-temperature superconductors is
very unusual.
Indeed, one observes that the isotope coefficient has a {\it minimum}
value at optimal doping ({\it highest} $T_c$). The explanation
of a small $\alpha$ for oxygen substitution coinciding with a high
$T_c$ is likely to be related to the contribution of the oxygen modes
to the pairing.

The main conclusion of this section is that the Coulomb
interaction and its logarithmic weakening leads to the deviation
of the isotope coefficient from the value $\alpha = 0.5$ and to the
non-universality of $\alpha$.

\subsection{Band Structure Effects; Transition Metals}\label{sssec:band}

Up to now we have only considered the effect on $T_c$ induced by
the isotopic shift of phonon frequencies. Furthermore, we have
considered the case where the phonon frequency appears explicitely
in the expression for $T_c$ (as prefactor and in $\mu^{\star}$;
see Eqs.~(\ref{TcBCS}), (\ref{mcmillan}) and (\ref{Tckresin})).
Band structure effects have first been considered to explain the
discrepancies observed between the results following from the
strong-coupling McMillan equation (Eq.~(\ref{mcmillan})) and the IC
measured in transition metals.
Several other generalizations of the basic model have been
considered, especially in the context of high-temperature
superconductors.

Let us first describe the situation encountered in transition metals.
As has been shown in Refs.~\cite{swihart,morel}, the two-square
model for which (\ref{mustar}) is derived leads to values of the IC
that are only $10-30\%$ lower than $0.5$. However,
the deviation is much stronger for some transition metals, leading
even to a vanishing coefficient for Ruthenium. As shown
in Ref.~\cite{garland} it is necessary to take into account the
band structure of transition metals. The presence of a ``metallic''
$s$-band and a narrow $d$-band, and the
associated peaked structure of the electronic density of states
(DOS) has an impact on the value of $\mu^{\star}$. Furthermore,
$E_F$ is much smaller in transition metals than in
non-transition metals and leads thus to a smaller effective
screening of the Coulomb interaction. These facts
have been included in Ref.~\cite{garland} by modifying the two
square-well model used to derive Eq.~(\ref{mustar}).
The values obtained from the modified expression of $\alpha$ (see
Eq.~(9) in Ref.~\cite{garland}) are in good semi-quantitative
agreement with the experimental results.

One should add that the two-band structure is important even in dirty
transition-metal superconductors, although the superconducting state
is characterized by only one gap in this case (see
Ref.~\cite{lynton} for further discussions on transition metals).

The example of transition metals shows that a quantitative
description of the IC requires, among other things, a precise
knowledge of the band structure. This has also been suggested in
Ref.~\cite{fowler,meservey} to explain the large inverse isotope
effect observed in Uranium. In this context, it would be interesting
to determine the uranium isotope effect in heavy-fermion systems
(one should note, however, that there is only one report of the
isotope effect in uranium\cite{fowler}).

Another effect of the band structure arises if one assumes that
the electronic density of states varies strongly on a scale given by
$\Omega$, the BCS energy cutoff (e.g., in the presence of a van Hove
singularity). To obtain the BCS expression (\ref{TcBCS}) we assumed
that the electronic density of states (DOS) is constant in
the energy intervall $[-\Omega,\Omega]$ around the Fermi energy.
$\lambda$ is then given by $N(E_F)V$, where $N$ is the DOS
at the Fermi level and $V$ is the attractive part of the
electron-electron effective interaction. In a more general case,
however, one has to consider the energy dependence of the DOS.
Though a purely electronic parameter, the energy dependence of
the DOS appears to influence the isotope effect\cite{schachinger}.
One
can understand this effect qualitatively within a crude extension of
the BCS model. Instead of taking the DOS at the Fermi level, one
replaces $N(E_F)$ by $<N(\varepsilon)>_{\Omega}$, the average value
of the DOS $N(\varepsilon)$ over the interval $[-\Omega,\Omega]$
around $E_F$. Obviously, this average depends on the cutoff
energy $\Omega$ of the pairing interaction. Isotopic substitution
modifies $\Omega$ which affects $<N(\varepsilon)>_{\Omega}$, and thus
$\lambda=<N>_{\Omega}V$ and $T_c$.

A more general analysis of this effect within Eliashberg's theory
was given in Ref.~\cite{schachinger}. They find that whereas the IC
can reach values above $0.5$, its minimal value obtained for
reasonable choices of the parameters never reaches the
small values ($\sim 0.02$) found in optimally doped high-$T_c$
compounds. However, more recent studies of the influence of van Hove
singularities on the isotope coefficient show that such small
values can be obtained, because in this scenario the cutoff energy
for the effective interaction between charge-carriers is given by
$min(E_F-E_{vH}; \Omega)$ (where $E_{vH}$ is the energy of the
van Hove singularity and $\Omega$ is a characteristic phonon energy)
\cite{labbe,abrikosov,hocquet}. If the cutoff energy is
given by $E_F - E_{vH}$ and is thus electronic in origin, $T_c$
displays no shift upon isotopic substitution. Taking
into account the Coulomb repulsion it has been shown\cite{hocquet}
that one can even obtain a negative IC.

Another way to extend the results presented earlier is to consider
an anisotropic Eliashberg coupling function
$\alpha^2F({\bf q},\omega)$. Generally it is assumed that the system
is isotropic and the coupling function can be averaged over the
Fermi surface, leading to a {\bf q}-independent Eliashberg function.
If the system is strongly anisotropic (as is the case of
high-$T_c$ superconductors), the average may not lead to an accurate
description of the situation.
The isotope effect has been studied for different {\bf q}-dependent
form factors entering $\alpha^2F$ in Ref.~\cite{dahm}. They show
that a small isotope coefficient as observed in
YBa$_2$Cu$_3$O$_{7-\delta}$ can be obtained for anisotropic
systems. Nevertheless, to obtain high critical
temperatures at the same time, they are forced to introduced another,
electronic pairing mechanism. Such a situation is further discussed
below (see Sec.~\ref{ssec:nonph}).

Note that the theories presented in this section give only a
qualitative
picture of the situation encountered in high-temperature
superconductors and do presently not account for the isotope effect
observed in various systems (see also
Refs.~\cite{schachinger,crespi}).

\subsection{Polyatomic Systems}\label{ssec:poly}

Until now we focused mainly on the study of monoatomic systems in
which the attractive interaction leading to
the formation of pairs is mediated by phonons.
All previous effects are naturally also encountered in polyatomic
systems. However, the presence of two or more elements in the
composition of a superconductor has several direct consequences for
the isotope effect that we summarize in the following.
The first consequence of a polyatomic system is that the
characteristic frequency $\Omega$ which determines the value of $T_c$
depends on the mass
of the different ions involved $\Omega = \Omega(M_1,M_2,\cdots)$.
Obviously, the dependence $\Omega\sim M_r^{-\alpha_r}$ for element
$r=1,2,...$ must not be equal to $\alpha_r = 0.5$.
It is thus not surprising if the partial isotope coefficient
(obtained by substituting one type of ions for its isotope; see
Eq.~(\ref{ICdelta})) differs from the textbook value $0.5$.

Let us consider the example of a cubic lattice with alternating
masses $M_1$ and $M_2$\cite{maradudin,kresinlivre}. The acoustic
and optical branches can be calculated analytically\cite{maradudin}.
From this one obtains the following partial isotope coefficients:
\begin{eqnarray}
\alpha_1 = \frac{1}{2\Omega^2 M_1} \left\{ \bar{K}\pm
\frac{\bar{K}^2 M_{12}^-
+ 2\bar{K}_L/M_2}{\left(\Omega \mp
\bar{K}M_{12}^+\right)}\right\} \qquad,\\
\alpha_2 = \frac{1}{2\Omega^2 M_2} \left\{ \bar{K}\pm
\frac{-\bar{K}^2M_{12}^-
+ 2\bar{K}_L/M_1}{\left(\Omega \mp
\bar{K}M_{12}^+\right)}\right\} \qquad,
\end{eqnarray}
where $M_{12}^- = M_1^{-1} - M_2^{-1}$, $M_{12}^+ = M_1^{-1} +
M_2^{-1}$. The upper sign stands for the Debye frequency and the
lower sign has to be considered when the characteristic frequency is
given by the optical phonon. $\bar{K} = \sum_{x,y,z} K_i$ is the
sum of the force constants and $\bar{K}_L = \sum_{x,y,z}
K_i\cos(q L)$ ($q=0$ for the optical branch and $q=\pi/L$ for
acoustic mode; $L$ is the lattice constant). One notes that the
total isotope coefficient
is given by $\alpha_{tot} = \alpha_1+\alpha_2 = 0.5$ whether one
takes the acoustical or the optical phonon as the characteristic
energy for the determination of $T_c$.

Given the previous result, one can ask if there exists a maximal
value of the IC. It is often stated in the literature
that $0.5$ is the maximal value that the isotope coefficient
can reach within a harmonic phonon, and electron-phonon induced
pairing model. Although to the knowledge of the authors none of the
systems studied so far seems to contradict this assertion, one
should note that there is no proof of this statement.
It was shown in Ref.~\cite{rainer} that for a polyatomic system
$\alpha_{tot} \equiv \sum_r \alpha_r = 0.5$ (defined in
Eq.~(\ref{ICTcpartial}))
{\it if} one assumes that $\mu^{\star} = 0$ and all masses of the
unit cell are subject to {\it the same} isotopic shift
(i.e., $\Delta M_r/M_r$ takes the same constant value for each
element $r$ of the system, see Eq.~(\ref{ICTcpartial})). These
assumptions hold approximately for certain systems as, e.g., the
Chevrel-phase Mo$_6$Se$_8$ material\cite{rainer}, but are certainly not valid for
example
in high-temperature superconductors. It is thus not clear if the
{\it total} isotope coefficient, Eq.~(\ref{ICTcpartial}), is indeed
always $0.5$. The
general proof of such a statement requires the knowledge of the
polarization vectors and their derivatives with respect to the
isotopic masses, both of which have to be calculated for each
specific system studied.

Another important remark concerns the value of the (very small) IC
for optimally doped high-$T_c$ superconductors. Experimentally one
measures generally the partial oxygen isotope effect (in some cases
also the Cu and Ba IE). Since the unit cell of a high-temperature
superconductor contains many different atoms, one expects values of
the partial IC that are significantly smaller than $0.5$.
A crude estimate can be given by observing that under the assumption
of a same contribution of each atom to the isotope coefficient (which
is certainly not the case as mentioned above) the value of the
isotope effect for oxygen in YBCO would be $\alpha \approx 0.5/N \approx
0.04$, where $N$ is the number of atoms per unit cells ($N=13$ in
the case of YBCO). Since there are $6-7$ oxygen atoms in the unit
cell, one should multiply $\alpha_{ox}$ by this number.

If, in addition to the multiatomic structure of the system, one
takes into account the effect of Coulomb interactions
and/or the fact that the charge-carriers may strongly couple to
certain phonons only one can obtain values of the order observed in
high-temperature oxides. The second possibility has for
example been studied in the context of the Chevrel-phase compound
Mo$_6$Se$_8$\cite{rainer} or organic superconductors\cite{auban}.
It has also been studied for high-$T_c$
superconductors in Ref.~\cite{ashauer,barbee}. They conclude that the
oxygen isotope effect can be well described for reasonable parameters
in the case of La$_{2-x}$Sr$_x$CuO$_4$, but that unphysical values
of the coupling have to be considered to describe the partial oxygen
isotope effect of YBa$_2$Cu$_3$O$_{7-\delta}$. A careful study of
the effect of Coulomb interactions on the IE of high-$T_c$
superconductors has not been done yet.

On thus concludes from the previous considerations that it is not
impossible to explain the small
values of the IC observed in conjuction with high $T_c$'s in
high-temperature materials, if one considers the combined effect of a
polyatomic system, the Coulomb interaction and the band structure.

\subsection{Anharmonicity}\label{ssec:anh}

Anharmonic effects play an important role in materials such as
PdH(D)\cite{stritzker,ganguly}, Mo$_6$Se$_8$\cite{rainer} or, according to
some theories\cite{crespi,galbaatar,muller,cyrot}, in
high-temperature superconductors.
Among other consequences the presence of anharmonicities affects the
value of the isotope coefficient. We present here two different
aspects of the anharmonic isotope effect: anharmonicity of the
characteristic phonon mode (for PdH) and volume effects (for
Molybdenum).

Quantities that are independent of ionic masses in the harmonic
approximation, can become mass dependent in the presence of
anharmonicity. We mention three such properties that are of
interest for the isotope effect in superconductors: the
lattice force constants $K_i$\cite{ganguly}, the unit-cell volume
\cite{jansen,nakajima} and the electron-phonon
coupling function $\lambda$\cite{crespi,galbaatar}.
In the first case, it was shown that the account of Coulomb
interactions alone (see above) cannot explain the inverse
($\alpha<0$) hydrogen
isotope effect observed in PdH (see Table 1). On the other hand, a
change of $20\%$
of the lattice force constants when replacing H for Deuterium (D)
in PdH was infered from neutron-scattering data\cite{ganguly}.
This results from large zero-point motion of H as compared to D.
Taking this fact into account in the Eliashberg formalism allows
one to obtain quantitative agreement with the
experiment\cite{ganguly}.

Another effect due to the presence of anharmonicity is the isotopic
volume effect\cite{jansen,nakajima}. This effect was discussed in the
context of Molybdene and is related to the difference in
zero-point motion of the two isotopes. In such a system one can
write the IC as $\alpha = \alpha_{BCS} + \alpha_{vol}$ with
\begin{eqnarray}
\alpha_{vol} = B \frac{M}{\Delta M}\frac{\Delta V}{V}
\frac{\Delta T_c}{\Delta P} \qquad,
\end{eqnarray}
where $B = -V \partial P/\partial V$ is the bulk modulus, $P$ is the
pressure and $\Delta V = V^{\star} - V$ is the volume difference
induced by isotopic substitution. In the case of Molybdenum the
volume isotope effect amounts to $\alpha_{vol} \approx 0.09$ and
accounts for $\sim 27\%$ of the total isotope coefficient, which is
not a negligible effect.

The presence of a volume isotope effect has also been suggested in
the case of PdH, C$_{60}$ materials as well as in Pb\cite{jansen}.

The other major effect of anharmonicity is the appearence of an
ionic-mass dependent electron-phonon coupling function
$\lambda$\cite{crespi,galbaatar,schlutter}. In this case not only
the prefactor $\Omega$ of Eq.~(\ref{TcBCS}), but also the exponent
depends on ionic masses. As stressed earlier in the context of band
structure effects, this fact can lead to a strong deviation from
$\alpha_{BCS} = 0.5$, even for monoatomic systems. An explicit
expression for the motion of the oxygen in a simple double-well
potential was obtained in Ref.~\cite{galbaatar} for high-$T_c$
superconductors.
The model was extended further in Refs.~\cite{galbaatar}b and
\cite{crespi} and it was shown that the isotope coefficient can
be much smaller than $0.5$ with moderate values of the coupling
constant or exceed $0.5$ for (very) strong couplings (the
function $\alpha(\lambda)$ goes through a minimum at $\lambda\sim
1$; see Ref.~\cite{galbaatar}). Regarding high-$T_c$ superconductors,
the anharmonic model could explain qualitatively the behaviour of the
isotope coefficient if one considers the fact that the coupling
function $\lambda$ also depends on doping
(see Ref.~\cite{morawitz1} for this dependency). Assuming that these
systems are strong-coupling superconductors, and that the coupling
decreases upon doping, one can describe qualitatively the behaviour
of $\alpha(T_c)$ for underdoped systems. However, it is difficult to
obtain quantitative agreement especially in the optimally doped
and strongly underdoped regime. In the first case the coupling has
to be intermediate to obtain small $\alpha$'s but then $T_c$ is also
small. In the second, strongly underdoped case, the IC can exceed
$0.5$, but the coupling has to be very strong. One way to solve
this problem is to assume that the superconducting pairing is
mediated by an additional, non-phononic channel (see
Ref.~\cite{schlutter} and below).

The influence of anharmonicity on the electron-phonon
coupling $\lambda$ and on the hopping parameter (for a tight-binding
type of lattice) has also been considered in the context of
$^{12}$C $\leftrightarrow^{13}$C isotope substitution in
$A_3$C$_{60}$ materials ($A=$Na, Ru; see, e.g.,
Ref.~\cite{ashcroft} and references therein). Experimentally the
isotope coefficient was shown to vary between $0.37$ to $1.4$,
depending on the isotopic substitution process (in the first case
each C$_{60}$ molecule contains an equal amount of $^{12}$C and
$^{13}$C isotopes whereas in the second case the system contains
C$_{60}$-molecules that are composed of either pure $^{12}$C or pure
$^{13}$C atoms) \cite{ashcroft}.

\subsection{Non-Phonon and Mixed Mechanisms}\label{ssec:nonph}

The superconducting transition can also be caused by a non-phonon
mechanism. Historically, the introduction of the electronic
mechanisms \cite{little} started the race for higher $T_c$'s.
The pairing can be provided by the exchange of excitations such as
excitons, plasmons, magnons etc... (see, e.g.,
Ref.~\cite{kresinlivre}).
In general, one can have a combined mechanism. An electronic channel
can provide for an additional contribution to the pairing, as is
the case, e.g., in a phonon-plasmon mechanism \cite{morawitz2}.
In the context of the isotope effect, the mixed mechanism may
provide for an explanation for the unusual occurence of high $T_c$'s
and small isotope coefficients $\alpha$ (see above).

Let us initially consider the case where only electronic excitations
mediate the pairing interaction and
the Eliashberg function $\alpha^2F$ can be approximated by a
single peak at energy $\Omega_e$, the characteristic electronic
energy. The theory considered in weak coupling
yields then a relation of the form $T_c\simeq
\Omega_e e^{-1/\lambda}$.
For an electronic mechanism $\Omega_e$ is independent of the
ionic mass. Therefore, the isotope coefficient of $T_c$ vanishes for
such cases.

If, on the other hand, the superconducting state is due to the
combination of phonon and a high-energy excitations
(as, e.g., plasmons \cite{morawitz2}) then one has to include both excitations in
$\alpha^2F$. In this case, the simplest model considers an
Eliashberg function
composed of two peaks, one at low energies ($\Omega_0$) for the
phonons and one at high energies ($\Omega_1$) for the electronic
mechanism. For such a model, one
obtains the following expression for $T_c$\cite{geilikman}:
\begin{eqnarray}
T_c = 1.14 \Omega_0^{f_0}\Omega_{1}^{f_{1}}
\exp{\left(-\frac{h(\rho_i,\Omega_i)}{\rho_0 + \rho_1}\right)}
\end{eqnarray}
where $h$ is a slow varying function of $\rho_i$ and $\Omega_i$,
$f_i = \rho_i/(\rho_0 + \rho_1)$ and
$\rho_i = \lambda_i/(1+\lambda_0+\lambda_1)$ ($i=1,2$). In weak
coupling, $h\approx 1$ and the exponent is independent of the ionic
masses. The resulting isotope coefficient for this mixed mechanism
takes then the form:
\begin{eqnarray}
\alpha = \frac{f_0}{2} = \frac{1}{2}
\left\{ 1 - \frac{\lambda_1}{\lambda_0 + \lambda_1}\right\} \qquad .
\end{eqnarray}
If only the phonon mechanism is active $\lambda_1 = 0$ and
one recovers the BCS value $0.5$. Otherwise, the presence of the
non-phononic mechanism reduces the value of the isotope coefficient,
while enhancing $T_c$.
For $\lambda_0\ll \lambda_1$ one has $\alpha \approx 0$. A joint
mechanism would thus allow one to explain the small values of the IC
for optimally doped high-$T_c$ superconductors since the coupling
to non-phononic degrees of freedom would provide for high $T_c$'s
while it would reduce, together with the Coulomb repulsive term,
the value of the IC.

Another way to include non-phononic contributions to the pairing
mechanism, is to consider a negative effective Coulomb term
$\mu^{\star}$ in the Eliashberg
theory\cite{kresinlivre,wolfmustar,ashauer}. The
pseudo-potential $\mu^{\star}$ can thus be
seen as an effective attractive pairing potential contribution and
thus supports superconductivity. Such a negative
$\mu^{\star}$ is for example obtained in the plasmon model
\cite{morawitz2,kresinlivre,wolfmustar}. One notes that the small
values of the partial
isotope coefficient in optimally doped high-$T_c$ materials can
easily be explained within such a model.

Several other non-phononic as well as combined mechanisms have been
proposed \cite{ashauer,schlutter,das,marsiglio} that also lead to a
reduction of the isotope effect. We refer to the literature for the
details.

\subsection{Isotope Effect of Properties other than $T_c$}
\label{sssec:other}

All previous considerations were concerned with the isotope
effect of the superconducting critical temperature $T_c$. One can
ask if there are other properties displaying an isotope effect in
conventional superconductors. Naturally, every quantity depending
directly on phonon frequencies will display such an effect.
Let us consider here a property that will be studied further in the
next sections, namely the penetration depth of a magnetic field
$\delta$. In the weak-coupling London limit the penetration depth is
given by the well-known relation
\begin{equation}\label{london}
\delta^{2} = \frac{m c^2}{4\pi n_s e^2}
= \frac{m c^2}{4\pi n\varphi(T/T_c) e^2} \qquad .
\end{equation}
where $m$ is the effective mass. $n_s$ is the superconducting
density of charge carriers, related to the normal density $n$
through $n_s = n\varphi(T/T_c)$. The function $\varphi(T/T_c)$ is a
universal function of $T/T_c$. For example,
$\varphi \simeq 1-(T/T_c)^4$ near $T_c$, whereas $\varphi \simeq 1$
near $T=0$ (in the absence of magnetic impurities; their
influence is discussed in section \ref{sec:MI}).

Eq.~(\ref{london}) does not depend explicitely
on phonon frequencies. Nevertheless, it can display an isotopic
dependence through $T_c$. In fact, this dependency is common to
all superconducting properties within the BCS theory. Indeed, all
major quantities such as heat capacity, penetration depth, critical
field, thermal conductivity etc... can be expressed as universal
functions of $T_c$. As a result, all these
quantities display a trivial isotopic shift, caused by the isotopic
shift in $T_c$. The value of the shift is growing as one approaches
$T_c$. Furthermore, the shift vanishes for $T\rightarrow 0$.
As will be shown in section \ref{sec:MI} and following the situation
is very
different in high-temperature superconductors and manganites.
For example, the influence of non-adiabaticity on charge-transfer processes
results in an unconventional ionic mass
dependence of the charge-carrier concentration $n = n(M)$, which,
according to Eq.~(\ref{london}),
leads to a new isotope effect (see sec.~\ref{sec:NA}).\\

So far, we have presented various theories explaining the deviation
of the measured isotope coefficient in conventional superconductors
from the value $\alpha=0.5$ derived within the BCS model. The
remainder of this review is devoted to the study of three factors
that are not related to the pairing
interaction but affect the isotope coefficient: magnetic impurities,
proximity contacts, and non-adiabatic charge-transfer processes.
We show that they can strongly modify the value of the isotope
coefficient and can even induce an isotopic shift of
superconducting properties such as the penetration depth $\delta$.
In the last part of this review, we then apply the theory to the
case of the oxygen isotopic substitution in high-temperature
superconductors.

\section{MAGNETIC IMPURITIES AND\\THE ISOTOPE EFFECT}
\label{sec:MI}

The presence of magnetic impurities strongly affects various
properties of a superconductor. It has been
shown\cite{AG,skalski} that because of magnetic
impurity spin-flip scattering processes, Cooper pairs are broken
and thus removed from the superconducting condensate. Several
properties are affected by the pair-breaking effect. Magnetic
scattering leads to a decrease of the critical temperature
$T_c$, the energy gap, the jump in the specific heat at $T_c$, etc...
At some critical impurity concentration $n_M = n_{M,cr}$ superconductivity
is totally suppressed. Moreover, there exists an impurity concentration
$n_M = n_{M,g} < n_{M,cr}$ (for conventional superconductors
$n_{M,g} = 0.9 n_{M,cr}$, see Ref.~\cite{AG,skalski}) beyond which
the superconducting state is gapless. Pair-breaking also leads to an
increase
of the penetration depth $\delta$. We show in the following sections
that the presence of magnetic impurities leads to a
change of the isotope coefficient of $T_c$ and $\delta$.

\subsection{The Critical Temperature}

The change of $T_c$ induced by magnetic impurities is described
in the weak-coupling regime by the Abrikosov-Gor'kov
equation\cite{AG}:
\begin{eqnarray}\label{TcAG}
\ln\left(\frac{T_{c0}}{T_c}\right)
= \psi\left(\frac{1}{2} + \gamma_s \right)
- \psi\left(\frac{1}{2}\right) \qquad ,
\end{eqnarray}
where $\gamma_s = \Gamma_s/2\pi T_c$ and $T_c$ ($T_{c0}$) is the
superconducting critical temperature in the presence (absence) of
magnetic impurities. $\Gamma_s = \tilde{\Gamma}_s n_M$ is the
spin-flip scattering amplitude and is proportional to the magnetic
impurity concentration $n_M$ ($\tilde{\Gamma}_s$ is a constant).

It is easy to derive a relation between the isotope coefficient
$\alpha_0$ in the absence of magnetic
impurities [$\alpha_0 = -(M/\Delta M)(\Delta T_{c0}/T_{c0})$,
Eq.~(\ref{ICTc})] and its value in the presence of magnetic
impurities [$\alpha = -(M/\Delta M)(\Delta T_c/T_c)$]. From
Eqs.~(\ref{ICTc}) and (\ref{TcAG}) one
obtains\cite{carbotte,KBWO,BKW1,singh}:
\begin{eqnarray}\label{ICTcMI}
\alpha_{m} =
\frac{\alpha_0}{1 - \psi^{\prime}(\gamma_s + 1/2)\gamma_s} \qquad ,
\end{eqnarray}
where $\psi^{\prime}$ is the derivative of the psi function and is a
positive monotonous decreasing function of $\gamma_s$.
The fact that $\alpha_0$ and
$\alpha$ are not identical relies on the essential feature that the
relation between $T_{c0}$ and $T_c$ (Eq.~(\ref{TcAG})) is non-linear.
It is interesting to note that magnetic scattering decreases $T_c$
(reduction of the condensate) but {\it increases} the
value of the IC ($\alpha_m > \alpha_0$; see  Eq.~(\ref{ICTcMI})).

An important consequence of Eqs.~(\ref{TcAG}) and (\ref{ICTcMI}) is
that one can write the IC as a universal function $\alpha(T_c)$. This relation
does not contain any adjustable parameter, but depends solely on the
measurable quantities $\alpha_0$ and $T_c$. The universal curve
is shown in Fig.~2.
\unitlength1cm
\begin{figure}[htb]
\begin{center}
\input{Imp_UnivZnN.ps}
\end{center}
{\small {\bf Figure 2.} Isotope coefficient of $T_c$ in the presence of magnetic impurities. Solid line: Universal dependence $\alpha(T_c)$ (normalized to $\alpha_0$; see Ref.~\cite{KBWO}); Points: Isotope effect for YBa$_2$(Cu$_{1-x}$Zn$_x$)$_3$O$_7$ obtained with various experimental techniques and normalized to $\alpha_0 = 0.025$ (from Ref.~\cite{soerensen,KBWO}).}
\end{figure}
Note also that $\alpha_m$ can, in principle, exceed the value
$\alpha_{BCS} = 0.5$. Thus, the observation of large values of
$\alpha$ are not necessarily related to the pairing mechanism.

\subsection{Zn-doped YBa$_2$Cu$_3$O$_{7-\delta}$}

Zn substitution for Cu in the CuO$_2$ planes of YBCO has attracted a lot
of interest, since it leads to a drastic decrease in $T_c$ \cite{zagoulev}
and a strong increase of the penetration depth \cite{panagopoulos}
(superconductivity is destroyed with $\sim 10\%$ of Zn). In addition,
this decrease of $T_c$ is accompanied by an increase of the isotope
coefficient (see Ref.~\cite{zagoulev,soerensen} and Fig.~2).
We think that the peculiar behaviour of Zn doping is related to
the pair-breaking effect. It has been established by several
methods (see Refs.~\cite{zagoulev,panagopoulos}) that Zn substitution leads to
the formation of local magnetic moments ($\sim 0.63 \mu_B/$Zn, where
$\mu_B$ is the Bohr magneton) in the vicinity of Zn.

Fig.~2 displays the experimental results obtained for the
oxygen isotope coefficient of Zn-doped YBCO\cite{soerensen,KBWO} (normalized
to $\alpha_0 \simeq 0.025$\cite{zech}). One can see from this figure
that the agreement between the theoretical dependence $\alpha(T_c)$
and the experiment is very good. Note that the uncertainty in the data
is growing as $T_c\rightarrow 0$ (see Ref.~\cite{soerensen}).

\subsection{The Penetration Depth}\label{sssec:deltaMI}

The isotope effect of the penetration depth is more complicated than
the isotope effect of $T_c$, since
it appears to be temperature dependent. Here we present the main
results near $T_c$ and at $T=0$ and refer the reader to
Ref.~\cite{BKW1,BKW2} for details. Near $T_c$, the penetration depth is
given in second order of the superconducting order parameter
$\Delta\equiv \Delta(T,\Gamma_s)$ by\cite{skalski}:
\begin{eqnarray}\label{delMI}
\delta^{-2} = \sigma \frac{\Delta^2}{T_c}
\zeta(2,\gamma_s + \frac{1}{2}) \qquad ,
\end{eqnarray}
where $\sigma = 4\sigma_N/c$ ($\sigma_N$ is the normal state
conductivity), $\zeta(z,q) = \sum_{n\ge 0} 1/(n+q)^z$ and $\gamma_s$
was defined in Eq.~(\ref{TcAG}). One can calculate the isotope effect
from Eqs.~(\ref{ICdelta}), (\ref{delMI}) and the analytical
expression for $\Delta$ near $T_c$ (see Refs.~\cite{skalski,BKW1}):
\begin{eqnarray}\label{gapMI}
\Delta^2 = 2\Gamma_s^2(1-\tau)
\frac{1 - \bar{\zeta}_2 + (1-\tau)\left[ \frac{1}{2} -\bar{\zeta}_2 + \bar{\zeta}_3 \right]}{\bar{\zeta}_3 - \bar{\zeta}_4 }
\equiv 2\Gamma_s^2 \frac{N_1}{D_1}
\end{eqnarray}
with $\bar{\zeta}_z = \gamma_s^{z-1} \zeta(z,\gamma_s + 1/2)$,
$z=1,2,\ldots$ and $\tau = T/T_c$.

As mentioned in Sec.~\ref{sec:HIST},
$\delta$ experiences a trivial BCS isotope effect near $T_c$. Since
we are only
interested in the unconventional IE resulting from the presence of
magnetic impurities, we substract the BCS isotope effect by
calculating the isotope coefficient $\tilde{\beta}_m$ of
$\delta(T,\Gamma_s)/\delta(T,0)$. Near $T_c$ this coefficient can
be written as $\tilde{\beta}_m = \beta_m - \beta_0$, where $\beta_m$
($\beta_0$) is the isotope coefficient of $\delta$ in the presence
(absence) of magnetic impurities. One can show\cite{BKW2} that
(near $T_c$) $\tilde{\beta}_m$ can be rewritten in the form:
\begin{eqnarray}\label{ICdeltaMI}
\tilde{\beta}_m = (R_1 - R_0)\alpha_m \qquad ,
\end{eqnarray}
where
\begin{eqnarray}
\label{R1}
R_1 &=& -\frac{1}{2}f_1 = - \frac{1}{2}\left( \frac{N_2}{N_1} - \frac{D_2}{D_1}
+ 2\frac{\bar{\zeta}_3}{\bar{\zeta}_2} - 1 \right)  \qquad ,\\
\label{R0}
R_0 &=& - \frac{1}{2}\frac{\alpha_0}{\alpha_m}f_0
= -\frac{1}{2}\left[ 1 - \psi^{\prime}(\gamma_s + 1/2)\gamma_s \right]f_0
\qquad ,
\end{eqnarray}
and $f_0 = (3-\tau^2)/(1-\tau)(3-\tau)$.
The functions $N_1$, $D_1$ are defined in Eq.~(\ref{gapMI}) and
\begin{eqnarray*}
N_2 &=& 3(1-\tau)^2
\left[ \bar{\zeta}_2 - 2\bar{\zeta}_3 + \bar{\zeta}_4  \right]
+ \tau(2-\tau) - \bar{\zeta}_2\\
D_2 &=& 2\left(3\bar{\zeta}_4 - \bar{\zeta}_3 - 2\bar{\zeta}_5\right)
\qquad .
\end{eqnarray*}
This relation is valid near $T_c$
(where $\Delta$ is small) and for impurity concentrations such
that $\Delta T_c/T_c \ll 1$.

One notes first that the isotope coefficient of $\delta$ is
proportional to the IC of $T_c$. 
As for the isotope coefficient of $T_c$ (Eq.~(\ref{ICTcMI})) all
quantities can either be obtained from experiment (e.g., $\alpha_0$,
$T_{c0}$ and $T_c$) or calculated self-consistently using
Eqs.~(\ref{TcAG}) and (\ref{ICTcMI}). There is thus no free
parameter in the determination of $\tilde{\beta}_m$.
\unitlength1cm
\begin{figure}[htb]
\begin{center}
\input{CIN_T.ps}
\end{center}
{\small {\bf Figure 3.}  Isotope coefficient $\tilde{\beta}_{m}$ near $T_c$ (normalized to $\alpha_0$) as a function of $T_c$ (i.e.~of magnetic impurity concentration) for $T/T_c = 0.75$ (solid line), $0.85$ (dashed) and $0.95$ (dotted). Multiplying by $\alpha_0 = 0.025$ gives the IC expected for YBa$_2$(Cu$_{1-x}$Zn$_x$)$_3$O$_{7-\delta}$.}
\end{figure}
Fig.~3 displays the IC $\tilde{\beta}_{m}$ (normalized to $\alpha_0$)
as a function of $T_c$ (i.e.~on the
concentration $n_M$) for fixed values of the temperature. Setting
$\alpha_0 = 0.025$, one obtains the relation $\alpha(T_c)$ expected
for YBCZnO.
It would be interesting to measure the isotope effect of $\delta$
near $T_c$ in this system, to verify our predictions.

An interesting property of Eq.~(\ref{ICdeltaMI}) that is shown in
Fig.~3 is the fact that the isotope coefficient of $\delta$ is
temperature dependent; an unusual feature in the context of
the isotope effect. We will see in Sec.~\ref{sec:prox} that a
temperature dependent IC is also observed for a proximity system.

As can be seen by comparing Figs.~2 and 3, the qualitative behaviour
of $\tilde{\beta}_m(T_c)$ is similar to $\alpha_{m}(T_c)$
[Eq.~(\ref{ICTcMI}) with $\alpha_0 = \alpha_{ph}$] but with opposite
sign. Furthermore, in the absence of magnetic impurities one has
$\tilde{\beta}_{m} = 0$, whereas $\alpha_{m}(T_c) = \alpha_{ph}$.

Note finally, that since $R_1-R_0 <0$, the IC of the penetration
depth is
always negatif when magnetic impurities are added to the system. This
conclusion might not hold in certain cases if the non-adiabatic
channel is also included (then one has $\beta_{tot} = \beta_m -
\beta_0 + \beta_{na}$ and $\beta_{na}$ is positive, see
Sec.~\ref{sec:NA}).\\

All previous considerations have been done near $T_c$. At $T=0$,
the BCS contribution arising from the $T_c$ dependency of the
superconducting charge-carrier density $n_s$ (see
Eq.~(\ref{london})) vanishes, because $\varphi(T=0) = 1$. The
only contribution to the IE is thus due to magnetic impurities.
In the framework of the Abrikosov-Gor'kov theory one can write the
isotope coefficient at $T=0$ in the form:
\begin{eqnarray}\label{ICdeltaMI0}
\beta_m = R_0 \alpha_0 \qquad ,
\end{eqnarray}
where $R_0$ is given in appendix. Note that $R_0$ is a
negative function of $\Gamma_2$ (the
direct scattering amplitude) and $\Gamma_s = \Gamma_1 - \Gamma_2$
(the spin-flip, or exchange scattering amplitude) as defined by
Abrikosov and Gor'kov\cite{AG}. The IC of $\delta$ is thus negative
both near $T_c$ and
at $T=0$. This has to be seen in contrast to the IC of $T_c$ which
is always positive (see Fig.~2).
\unitlength1cm
\begin{figure}[htb]
\begin{center}
\input{CIN_0.ps}
\end{center}
{\small {\bf Figure 4.} Isotope coefficient of the penetration depth at $T=0$ in the presence of magnetic impurities (normalized to $\alpha_0$). There is no free parameter in the relation $\beta_m(T_c)$. Solid and dashed lines are for $\Gamma_2/\Gamma_s = 10$ and $50$ respectively (see Ref.~\cite{BKW2}).}
\end{figure}
Fig.~4 shows the universal relation $\beta(T_c)$ for two values of
$\Gamma_2$. One remembers that $T_c$ is determined by the magnetic
impurity concentration.

There are two major differences
between the results obtained near $T_c$ (Fig.~3) and at $T=0$
(Fig.~4). First, the direct
scattering amplitude $\Gamma_2$ appears only at $T=0$. Secondly,
the IC near $T_c$ is proportional to $\alpha_m$ given by
Eq.~(\ref{ICTcMI}),
whereas the IC at $T=0$ is function of $\alpha_0$, the IC of $T_{c0}$
in the absence of magnetic impurities. This difference is due to the
fact that $\varphi = 1$ at $T=0$  (see Eq.~(\ref{london})) and the
penetration depth does consequently not
depend on $T_c$ (the critical temperature in the presence of
magnetic impurities).

As shown in Fig.~2, the IC of $T_c$ fits the data of Zn-doped YBCO.
On the other hand, there are no experimental data for conventional
superconductors. Furthermore, the magnetic impurity contribution
to the IE of $\delta$ shown in Fig.~3 has never been measured. It
would thus be interesting to perform measurements of these effects,
especially since they can be described by universal relations.

\section{ISOTOPE EFFECT IN A\\PROXIMITY SYSTEM}\label{sec:prox}

In this section we consider another case, a proximity
system, in which a factor not related to lattice dynamics induces
a change of the isotope coefficient of $T_c$ and $\delta$. Consider
an $S-N$ sandwich where $S$ ($N$) is a superconducting (normal)
film. The value of the critical temperature $T_c$ and the penetration
depth $\delta$ differ substantially from the values $T_{c0}$ and $\delta_0$
of the isolated superconductor $S$\cite{kresin2,kresin3}. We show in
the following that the proximity effect also influences the isotope
coefficient of $T_c$ and $\delta$.

\subsection{The Critical Temperature}

Let us consider a proximity system composed of a weak-coupling
superconductor $S$ of thickness $L_S$ and a metal or a semiconductor
$N$ of thickness $L_N$ (e.g.~Nb-Ag). In the framework of the
McMillan tunneling model\cite{mcmillan}, which can be used when
$\delta < L_N \ll \xi_N$ ($\xi_N = hv_{F;N}/2\pi T$ is the coherence
length of the $N$-film
as defined in Ref.~\cite{clarke}), the critical temperature $T_c$ of
the $S-N$ system is related to the critical temperature $T_{c0}$ of
$S$ by the relation\cite{kresin2}:
\begin{eqnarray}\label{Tcprox}
T_c = T_{c0}\left( \frac{\pi T_{c0}}{2\gamma u} \right)^{\rho}\, , \,
\rho = \frac{\nu_NL_N}{\nu_SL_S} \qquad ,
\end{eqnarray}
where $\gamma\simeq 0.577$ is Euler's constant.
The value of $u$ is determined by the interplay of the McMillan
tunneling parameter $\Gamma = \Gamma_{SN} + \Gamma_{NS}$ and the
average phonon frequency $\Omega$. For an almost ideal $S-N$
contact ($\Gamma \gg \Omega$) one has $u\simeq \Omega$. In the
opposite limit, when $\Gamma \ll \Omega$ one obtains
$u \simeq \Gamma$ (see Ref.~\cite{kresin2}).

Let us first consider the case $\Gamma \gg \Omega$.
In the BCS model, one has $T_{c0}\propto \Omega$. Thus, according to
Eq.~(\ref{Tcprox}) $T_c$ and $T_{c0}$ have the same dependency
on the ionic mass in this limit (because $T_{c0}/u$ is then
independent of  $\Omega$). This results in the
simple relation $\alpha_{prox} = \alpha_0$ for the IC of $T_c$
($\alpha_0$ is the isotope coefficient of $T_c$ for the isolated
$S$-film).

The opposite situation where $\Gamma \ll \Omega$ is
more interesting since $u=\Gamma$ is independent of ionic masses.
Using Eqs.~(\ref{ICTc}) and (\ref{Tcprox}) one obtains for the
isotope coefficient of $T_c$ in the limit $\Gamma \ll \Omega$:
\begin{eqnarray}\label{ICTcprox}
\alpha_{prox} = \alpha_0\left( 1 + \frac{\nu_NL_N}{\nu_SL_S} \right)
\qquad.
\end{eqnarray}
One notes first that whereas the presence of a normal film on the
superconductor decreases $T_c$, the film induces an {\it increase} of
the isotope coefficient of $T_c$. The same is true for the presence
of magnetic impurities and for the non-adiabatic IE when
$\partial T_c/\partial n>0$ (see next section). The second interesting feature of
Eq.~(\ref{ICTcprox}) is that one can modify the value of
$\alpha_{prox}$ by changing the thicknesses of the films. For example,
if $\nu_N/\nu_S = 0.8$ and $L_N/L_S = 0.5$ then
$\alpha_{prox} = 0.28$.
By increasing the thickness of the normal film such that $L_N=L_S$,
one obtains $\alpha_{prox} = 0.36$.

We stress the fact that, as in the previous section on magnetic
impurities, the change of the IC is due to a factor not related to
lattice dynamics. Therefore, there is no reason for $\alpha_{prox}$
to be limited to values below $0.5$.

To the best of our knowledge the change in the IC caused by the
proximity effect has never been measured, even in conventional
superconductors. It would be interesting to carry out such
experiments in order to observe this phenomenon.

\subsection{The Penetration Depth}

The proximity effect also affects the shielding of a magnetic field.
The most dramatic effect of the normal
layer on the penetration depth is seen in the low-temperature regime
($T/T_c \le 0.3$). Although the penetration depth of a pure
conventional superconductor is only weakly temperature dependent in
this regime ($\delta^{-2} \sim \varphi = 1-T^4/T_c^4$ ) the
presence of the normal layer induces a temperature dependence
through the proximity effect. For the same $S-N$ proximity system as
considered above the penetration depth is given by\cite{kresin3}
\begin{eqnarray}\label{prox}
\delta^{-3} = a_N \Phi \qquad ,
\end{eqnarray}
where $a_N$ is a constant depending only on the material
properties of the normal film (it is ionic-mass independent) and
\begin{eqnarray}\label{phi}
\Phi = \pi T \sum_{n \ge 0} \frac{1}{x_n^2p_n^2 + 1}
\, , \qquad p_n = 1+ \varepsilon t \sqrt{x_n^2 + 1} \qquad .
\end{eqnarray}
$x_n = \omega_n/\epsilon_S(T)$ with $\omega_n = (2n+1)\pi T$ (the
Mastubara frequencies) and $\epsilon_S(T)$ is the superconducting
energy gap of $S$. In the weak-coupling limit considered here,
$\epsilon_S(0) = \varepsilon\pi T_{c0}$ with $\varepsilon \simeq
0.56$. The dimensionless parameter $t = \ell/S_0$
with $\ell = L_N/L_0$ and $S_0 = \Gamma/\pi T_{c0}$
($\Gamma \sim 1/L_0$ is the McMillan parameter\cite{mcmillan}).
$L_N$ and $L_0$ are the thickness of the normal film and some
arbitrary thickness, respectively (in the following we take
$L_0 = L_S$, the thickness of the superconducting film).
From Eqs.~(\ref{ICdelta}) and (\ref{prox}) one obtains the IC of
the penetration depth for the proximity system:
\begin{eqnarray}\label{ICdeltaProx}
\beta_{prox} = - \frac{2\alpha_0}{3\Phi}
\sum_{n>0} \frac{x_n^2p_n^2}{x_n^2p_n^2 + 1}
\left( 1 - \frac{\varepsilon t}{p_n\sqrt{x_n^2 + 1}} \right) \qquad,
\end{eqnarray}
where $\alpha_0$ is again the IC of $T_{c0}$ for the superconducting
film $S$ alone. This result has three interesting features. First,
as for the case of magnetic impurities, the IC of $T_c$ and $\delta$
have
opposite signs. This observation is valid as long as one does not
mix the different channels presented in this work (magnetic impurities,
proximity effect and non-adiabaticity). The addition of a
non-adiabatic contribution (next section) may lead to a different
conclusion. Secondly, one can see from Eq.~(\ref{ICdeltaProx}) that
$\beta_{prox}$ depends on the proximity parameter $t$. One can thus
modify the
IC either by changing the ratio $\ell = L_N/L_S$ or the McMillan
tunneling parameter $\Gamma$ (e.g.~by changing the quality of the
interface; see Ref.~\cite{BKW2}). Fig.~5 shows this dependence for
different temperatures.
\unitlength1cm
\begin{figure}[htb]
\begin{center}
\input{CIPrxN_t.ps}
\end{center}
{\small {\bf Figure 5.} Isotope coefficient (normalized to $\alpha_0$) of the penetration depth for a proximity system as a function of $t$, for $T/T_c=0.1$ (solid), $0.2$ (dashed), $0.3$ (dotted). See Ref.~\cite{BKW2}.}
\end{figure}
One notes that, contrary to the
value of $\alpha_{prox}$, the IC of the penetration depth decreases
with increasing ratio $\ell$. Finally, one notes that the IC
$\beta_{prox}$ is temperature-dependent. This unconventional feature
was also observed for the IC of $\delta$ in the presence of magnetic
impurities near $T_c$.
\unitlength1cm
\begin{figure}[htb]
\begin{center}
\input{CIPrxN_Temp.ps}
\end{center}
{\small {\bf Figure 6.} Isotope coeff.~$\beta_{prox}$ (normalized to $\alpha_0$) for a proximity system as a function of $T/T_c$ for $S_0=0.2$: $l=1$ (solid) $l=0.5$ (dashed) and $S_0=1$: $l=1$ (dash-dotted) $l=0.5$ (dotted). See Ref.~\cite{BKW1}.}
\end{figure}
Fig.~6 shows the temperature dependence for different values of the
parameters. The trend is similar to the case of magnetic impurities:
$|\beta_{prox}|$ increases with increasing $T$. Note, however, that
the two effects are calculated in very different temperature
ranges (near $T_c$ and near $T=0$). A complete description of this
effect can be found in Ref.~\cite{BKW1}.

\section{NON-ADIABATIC ISOTOPE EFFECT}\label{sec:NA}

The non-adiabatic isotope effect introduced in Ref.~\cite{KW1} was
used in Refs.~\cite{KW1,KBWO,BKW1,BKW2,KW2} to describe the
unusual
behavior of the isotope coefficient in several high-temperature
superconductors. The theory also allowed us to describe the large
isotopic shift of the ferromagnetic phase transition in manganites
\cite{KW2}. The concept of the non-adiabatic IE relies on the
fact that in the above-mentioned systems charge-transfer processes
involve a non-adiabatic channel. In other
words, the electronic ground state of the group of atoms over which
the charge-transfer takes place is degenerate, leading to a dynamic
Jahn-Teller (JT) effect ($E_{JT} \le \hbar \omega$, where $E_{JT}$
is the JT energy; see, e.g., Ref.~\cite{salem}).

The concept of the non-adiabatic isotope effect can be
understood by considering two examples: the apex oxygen in
high-$T_c$ oxides and the motion of charge-carriers in manganites.
Let us first consider high-$T_c$ materials. This system can be seen
as a stack of charge reservoirs (CuO chains) and conducting
subsystems (CuO$_2$ planes). Charges are transfered from the
reservoir to
the conducting layers through the apex oxygen that bridges the two
subsystems. Several experiments have shown\cite{sharma,mustre,haskel}
that the apex oxygen displays a non-adiabatic behaviour. The ion
oscillates between two close
positions that correspond to two configurational minima of the
potential energy surface. These two minima are due to the Jahn-Teller
effect since we are dealing with the crossing of electronic terms
(see Fig.~7).
\unitlength1cm
\begin{figure}[htb]
\begin{center}
\input{PES.ps}
\end{center}
{\small {\bf Figure 7.} Potential energy surface (PES) as a function of the configurational coordinate $R$. The crossing of electronic terms leads to a dynamical Jahn-Teller distortion (two minima of the PES).}
\end{figure}

The Jahn-Teller effect leads to a ``double-well'' type potential
(which should
not be confused with the double-well appearing when the crystal
is anharmonic; here we consider energy terms crossing and for each
term we use the harmonic approximation). This structure has a strong
impact on charge-transfer processes in these
materials. Indeed, the motion of charges from the chains to the planes occurs through
the apex oxygen. Thus, the charge-transfer process involves the
motion of the non-adiabatic ion. The immediate
consequence of this observation is that the density of
charge-carriers $n$ in the CuO$_2$-planes (the conducting subsystem)
depends on the mass of the non-adiabatic ions involved $n=n(M)$ (see
below).

The situation encountered in manganites is different from the
example just described. In these materials, there is no transfer
between a reservoir and a conducting subsytem. Instead, the motion
of charge-carriers occurs on the Mn-O-Mn complex that displays a
non-adiabatic behaviour\cite{sharma2}. As a consequence, the
electron hopping involves the motion of the ions and depends on
their mass. It was shown in Ref.~\cite{KW2} that this fact can
account for the unexpectedly low ferromagnetic critical temperature
$T_{c,f}$\cite{jonker} (which is determined by the hopping of the
electrons between Mn ions) as well as for the huge isotopic shift of
$T_{c,f}$\cite{zhao2}.

Note that although the charge-transfer processes are different in
high-temperature superconductors and in manganites, the formalism
described below and in Ref.~\cite{kresin} applies equally well to the
two systems. In the following we refer to the example of the
apex oxygen. The case of manganites can be mapped by replacing
``reservoir'' and ``conducting subsystem'' by ``Mn ions''.

It is important to realize that the dependence $n(M)$ is specific to
systems where the motion of charges occurs through non-adiabatic
ions. In metals, the charge-carriers are not affected by the motion
of the ions because of the validity of the adiabatic approximation.
We also stress the fact that the apex oxygen is given as a simple
example but, as will be shown below, the theory is not limited to
this case. Finally, one should note that a phenomenological theory
of the IC based on the assumption that $n=n(M)$ has been proposed in
Ref.~\cite{schneider}.

To demonstrate that in the presence of non-adiabatic charge-transfer
one has $n=n(M)$, one best uses the so-called diabatic
representation\cite{malley} (see also Ref.~\cite{kresin}). Here we
only recall the main steps of the calculation.
Let us assume that, because of the degeneracy of electronic states
(or Jahn-Teller crossing), the potential energy surface $E(R)$
($R$ is the
relevant configurational coordinate) of the group of non-adiabatic
ions (such as the apex O, in-plane Cu and O for high-$T_c$ materials
or Mn and O ions in manganites) has two close minima (see Fig.~7).
In the diabatic representation the total wave-function
$\Psi({\bf r,R},t)$ ({\bf r} is the electronic coordinate) is written
as a linear combination of the wave-functions in the two
crossing potential surfaces. The total wave-function cannot be
decomposed as a product of electronic and lattice wave-functions.
On the other hand, one can write the total wave-function as a sum of
symmetric
and antisymmetric terms. The energy splitting between these symmetric
and antisymmetric terms, which corresponds to the
inverse lifetime of oscillations between the configurations (minima
of the potential energy surface), has the form\cite{KW1,BKW1,KW2,kresin}:
\begin{equation}\label{h12}
H_{12} = <\Psi_1|H_e|\Psi_2> \simeq L_0 F_{12} \quad,
\end{equation}
where $H_{e}$ is the electronic part of the total Hamiltonian and
\begin{eqnarray}
\label{L0}
L_0 = L({\bf R_0}) &=&
\int d{\bf r}\, \psi^{\ast}_1({\bf r,R})H_e\psi_2({\bf r,R})\quad ,\\
\label{F12}
F_{12} &=& \int d{\bf R} \, \Phi^{\ast}_1({\bf R})\Phi_2({\bf R})
\quad.
\end{eqnarray}
The last equality in Eq.~(\ref{h12}) is obtained under the assumption
that the electronic wave-function $\psi_i$ ($i=1,2$) is a slowly
varying functions
of {\bf R}. $L_0$ can then be evaluated at ${\bf R_0}$, the
crossing of electronic terms, and taken out of the integral over
{\bf R}. It does not depend on ionic masses. On the other hand,
the important Franck-Condon factor (\ref{F12}) depends on the lattice
wave-functions $\Phi_i({\bf R})$, and thus on ionic masses.
Given the perturbation, Eq.~(\ref{h12}), one can calculate the
probability of finding the charge-carrier in the conducting layer
\cite{KW1,BKW1,KW2,kresin}. Note that non adiabaticity affects the
bandwidth of the reservoir and conducting bands rather than the
chemical potential (which is the same because of thermodynamic
equilibrium). Thus, in the case of YBCO, the plane
and chain bands have the same chemical potential, but different
widths. This affects the charge-carrier concentration in the
two subsystems.

Qualitatively, the
charge transfer can be visualized as a multi-step process. First,
the charge carrier can move from the reservoir to the group of ions
(e.g.~from the chains to the apical oxygen in YBCO).
Then, the non-adiabatic ions tunnel to the other
electronic term ($\Psi_1\rightarrow\Psi_2$). As a final step, the
charge carrier can hop to the conducting layer (from the apical
oxygen to the CuO$_2$ planes).
The crucial point of the theory is that as a result of this
multi-step process, the probability $P$ of finding the charge-carrier
in the conducting subsystem depends on Eq.~(\ref{h12}) and thus on
the Franck-Condon factor $F_{12}= F_{12}(M)$, Eq.~(\ref{F12}), that
depends on ionic masses. In other words
the charge-carrier concentration $n$, which is proportional to the
probability $P$, depends on the ionic mass $M$.
It is this unusual dependence $n=n(M)$, found in high-temperature
superconductors and manganites, that is responsible for the
unconventional isotope effect of $T_c$ and $\delta$.


\subsection{The Critical Temperature}

Given that $n=n(M)$, we can write the isotope coefficient of $T_c$ as
$\alpha = \alpha_{ph} + \alpha_{na}$, where
$\alpha_{ph} = (M/T_c)(\partial T_c/
\partial \Omega)(\partial \Omega/\partial M)$ is the usual (BCS)
phonon contribution ($\Omega$ is a characteristic phonon energy)
and the non-adiabatic contribution is given by:
\begin{eqnarray}\label{ICTcNA1}
\alpha_{na} =
\gamma \frac{n}{T_c}\frac{\partial T_c}{\partial n} \qquad ,
\end{eqnarray}
where the parameter $\gamma = - M/n (\partial n/\partial M)$ has a
weak logarithmic dependence on $M$ (see Ref.~\cite{KW1}).
This parameter should not be  confused with Euler's constant.
Eq.~(\ref{ICTcNA1}) shows that the IC of $T_c$ depends on the
doping of the conducting layer and on the relation $T_c(n)$. This
result was used in Refs.~\cite{KW1,KBWO} to analyse the IE
of high-temperature superconductors (see below).

\subsection{The Penetration Depth}

The concept of the non-adiabatic isotope effect relies on the fact
that the charge-carrier concentration depends on the ionic mass
because of the Jahn-Teller crossing of electronic terms. Obviously,
any quantity that depends on $n(M)$ will also display an
unconventional isotope effect. One such quantity is the
penetration depth of a magnetic field $\delta$ given in
Eq.~(\ref{london}). Note that the
relation $\delta^{-2}\sim n_s$ is also valid in the strong-coupling
case (see, e.g., Refs.~\cite{BKW1}).
From Eq.~(\ref{ICdelta}) one has
\begin{eqnarray}\label{betans}
\beta \equiv
-\frac{M}{\delta}\frac{\partial\delta}{\partial n_s}
\frac{\partial n_s}{\partial M}
= \frac{M}{2 n_s}\frac{\partial n_s}{\partial M} \qquad .
\end{eqnarray}
As mentioned in Sec.~\ref{sssec:other}, because of the relation
$n_s = n\varphi(T)$, one has to
distinguish two contributions to $\beta$. There is a usual (BCS)
phonon contribution, $\beta_{ph}$,
arising from the fact that $\varphi(T/T_c)$ depends on ionic mass
through the dependency of $T_c$ on the characteristic phonon
frequency. This BCS contribution was discussed in Sec.~\ref{sec:HIST}.
In the present paper we focus on the non-trivial
manifestation of isotopic substitution arising from the isotope
dependence of the charge-carrier concentration $n$. Such an effect
can even be observed in the low-temperature region where
$\varphi \simeq 1$.
From Eq.~(\ref{betans})  and the relation $n_s = n\varphi(T)$ it
follows that
\begin{eqnarray}\label{betaNA}
\beta = \beta_{ph} + \beta_{na} =
\frac{M}{2\varphi(T)}\frac{\partial\varphi(T)}{\partial M}
+
\frac{M}{2n}\frac{\partial n}{\partial M} \qquad ,
\end{eqnarray}
where $n(M)$ is the normal-state charge-carrier concentration.
Comparing Eq.~(\ref{ICTcNA1}) and the second term of the right hand
side of Eq.~(\ref{betaNA}) one infers that $\beta_{na} = -\gamma/2$
and thus establish a relation between the non-adiabatic isotope
coefficients of $T_c$ and $\delta$:
\begin{eqnarray}\label{ICTcNA2}
\alpha_{na}
= -2\beta_{na} \frac{n}{T_c}\frac{\partial T_c}{\partial n}
\qquad .
\end{eqnarray}
This result holds for London superconductors. The equation contains
only measurable quantities and can thus be verified experimentally.
 It is interesting to note that
$\beta_{na}$ and $\alpha_{na}$ have opposite signs when
$\partial T_c/\partial n > 0$ (which corresponds to the underdoped
region of high-$T_c$ materials).

Note, finally, that in the presence of magnetic impurities the
relation $n_s = n\varphi(T)$ remains valid, but $\varphi(T)$ depends
now on the direct scattering amplitude
$\Gamma_2$ defined in Sec.~\ref{sssec:deltaMI} (this results, e.g.,
in the inequality $n_s(T=0) < n$ in the gapless regime). As a
consequence, magnetic impurities affect the first term in
Eq.~(\ref{betaNA}) (that now depends on $\Gamma_2$), but leaves
$\beta_{na}$ and thus Eq.~(\ref{ICTcNA2}) unchanged.

\section{OXYGEN ISOTOPE EFFECT\\IN HIGH-$T_c$ MATERIALS}
\label{sec:HTSC}

In this section we briefly discuss experiments done on
high-temperature superconductors in the light of the theory exposed
in sections \ref{sec:MI} and \ref{sec:NA}. A detailed study of
the oxygen isotope effect in high-$T_c$ materials can be found in
Refs.~\cite{KW1,KBWO,BKW1,BKW2}.

The oxygen isotope effect of $T_c$ has been observed in a number
of experiments\cite{franck,zech,soerensen,morris,zhao}. Here we
discuss the results obtained on Pr-doped and oxygen-depleted
YBa$_2$Cu$_3$O$_{7-\delta}$ (YPrBCO and YBCO respectively).
We also discuss shortly the isotope effect of the penetration
depth observed in
La$_{2-x}$Sr$_x$CuO$_4$ (LSCO). The analysis of the isotope effect in
Zn-doped YBCO has already been presented in Sec.~\ref{sec:MI}.

\subsection{The Critical Temperature}

Let us first consider the isotope coefficient of $T_c$ for Pr-doped
YBCO (YPrBCO). Several experiments have established that the
Pr replaces Y which is located between the two CuO$_2$
planes of a unit cell. The doping affects YBCO mainly in two ways.
First, it was shown that because of the mixed-valence state of Pr,
holes are depleted from the CuO$_2$ planes (see,
e.g., Ref.~\cite{maple}). Secondly, as for Zn substitution,
praesodimium changes the magnetic impurity concentration of the
system\cite{soerensen,maple}. Regarding the IE, these experimental
facts imply that through the first effect, the non-adiabatic channel
is activated, whereas the second effect leads us to consider the
magnetic impurity channel for the calculation of $\alpha$.
The IE is therefore described by
$\alpha_t = \alpha_{ph} + \alpha_{m+na}$, where $\alpha_{ph} =
0.025$\cite{zech} is the phonon contribution, and $\alpha_{m+na}$
is defined by Eq.~(\ref{ICTcMI}) with $\alpha_0$ given by
Eq.~(\ref{ICTcNA1}). The resulting expression depends on two
parameters $\gamma$ and $\tilde{\Gamma}_s$, characterizing the
non-adiabatic and magnetic impurity channels, respectively. The
latter quantity has been extracted from
a fit to the relation $T_c(x)$ ($x$ describes the Pr
doping)\cite{maple} and is $\tilde{\Gamma}_s = 123$K. The parameter
$\gamma = 0.16$ is determined from the mean-square fit to the
experimental data. The result is shown in Fig.~8.
\unitlength1cm
\begin{figure}[htb]
\begin{center}
\input{PrTc.ps}
\end{center}
{\small {\bf Figure 8.} Dependence of the isotope coefficient $\alpha$ on $T_c$ for Y$_{1-x}$Pr$_x$Ba$_2$Cu$_3$O$_{7-\delta}$. Theory: solid line with $\gamma = 0.16$, $\tilde{\Gamma}_s = 123$K, $\alpha_0 = 0.025$; Experiment: dc magnetization, resistivity and ac susceptibility from Refs.~\cite{franck,soerensen}.}
\end{figure}
The theory is in
good agreement with the experimental data. Note that the negative
curvature at high $T_c$'s reflects the influence of the non-adiabatic
channel (magnetic impurities give a contribution with opposite
curvature), whereas the positive curvature seen at low $T_c$ is
due to the presence of magnetic impurities\cite{BKW1}.

Oxygen-depleted YBCO is very similar to the previous case in that
oxygen depletion both introduces magnetic impurities into the system
and removes holes from the CuO$_2$ planes. The same equations as
above can thus be used, however, with appropriate values of the
parameters. The result is shown in Fig.~9, together with the
experimental data from Ref.~\cite{zech}.
\unitlength1cm
\begin{figure}[htb]
\begin{center}
\input{OTc.ps}
\end{center}
{\small {\bf Figure 9.} Dependence of the isotope coefficient $\alpha$ on $T_c$ for YBa$_2$Cu$_3$O$_{6+x}$. Theory: solid line with $\gamma = 0.28$, $\tilde{\Gamma}_s = 15$K, $\alpha_0 = 0.025$; Experiment: points from Ref.~\cite{zech}.}
\end{figure}
Note the sharp drop observed near $T_c=60K$. This drop and the peak
above it are related to the presence of a plateau in the dependence
$T_c(n)$ near $T_c = 60K$ (see, e.g., Ref.~\cite{zech}).
Since Eq.~(\ref{ICTcNA1}) contains
the derivative $\partial T_c/\partial n$, $\alpha_{na}$ will be
nearly zero at $60$K. As one increases $T_c$,
the slope of $T_c(n)$ jumps to a large value, and decreases again
to zero as $T_c\rightarrow 90K$. This behavior of $T_c(n)$ explains
the maximum appearing in $\alpha_{na}(T_c)$ above $60$K. The peak
structure is thus specific to oxygen-depleted YBCO because neither
YPrBCO nor YBCZnO display such a plateau in the relation $T_c(n)$.
It would be interesting to verify this result, by measuring the
oxygen isotope effect between $60$ and $90$K.

The main difference between Pr-substituted (Fig.~8) and
oxygen-depleted YBCO (Fig.~9) lies in the fact that the
doping affects different ions of the system. Magnetic
impurities are introduced at different sites (on Y for the
first material and in the chains or the apical oxygen position for
the second system). Furthermore, non-adiabatic charge-transfer
processes involve
mainly Pr as well as Cu and O of the planes in YPrBCO, whereas they
involve mainly the apical O in oxygen-depleted YBCO (to a lower
extent, chain and in plane O and Cu are also involved).

Contrary to YPrBCO, it is not possible to extract the value of
$\tilde{\Gamma}_s$ (the contribution of magnetic impurities). The
best fit to the few data gives $\gamma=0.28$ and $\tilde{\Gamma}_s =
15$K. Two important points have to be noted concerning these
values. First, because of the limited data available, the
values may vary, although the order of magnitude will remain (see
also Ref.\cite{BKW2}). The other remark concerns the value of
$\tilde{\Gamma}_s$ for YPrBCO and YBCO. From these values one
concludes that oxygen depletion introduces a smaller amount of
magnetic moments than Pr doping. It would be interesting to perform
more isotope effect experiments on oxygen-depleted YBCO and to
determine the effective magnetic moment per depleted oxygen through
other means.

\subsection{The Penetration Depth}

The only experimental observation of the isotopic shift of the
penetration depth has been done on La$_{2-x}$Sr$_x$CuO$_4$
(LSCO)\cite{zhao}. There are no data available on YBCO-related
materials. Since only Pr-doped YBCO has one free parameter ($\gamma$)
we present the results of our theoretical calculations only in this
case. The oxygen-depleted case was studied as a function of the
two parameters $\gamma$ and $\tilde{\Gamma}_s$ in Ref.~\cite{BKW1}.

Let us begin with the case of LSCO. This case is simpler to study
than YBCO related materials, since no significant amount of magnetic
impurities has been detected in this material. Our analysis can
thus be carried out with Eqs.~(\ref{ICTcNA1}) or (\ref{ICTcNA2}).
The isotope coefficient has been measured for Sr concentrations
near $x\approx 0.11$ and $x\approx 0.15$. The first concentration
corresponds to the region where $T_c(x)$ experiences a small dip. The
origin of this dip is not well-established but is probably related
to electronic inhomogeneities and structural instability.
Since many factors affect $T_c$ at this Sr concentration, it is
difficult to interpret correctly the isotope effect of $T_c$ and
$\delta$.

The second concentration at which an isotope shift of $\delta$ has
been measured is at optimal doping ($T_c$ is maximal). The
experimental shift $\Delta \delta /\delta \simeq 2\%$ and
Eq.~(\ref{ICdelta}) allows one to determine the value of $\beta_{na}
\simeq 0.16$ and thus $\gamma \simeq 0.32$ through
Eq.~(\ref{ICTcNA2}). This result is in
good agreement with the values obtained for Pr-doped and O-depleted
YBCO (see Ref.~\cite{BKW2} for a discussion on LSCO). 

Let us now turn to the isotope coefficient of the penetration
depth for YBCO-related materials.
From the evaluation of the parameter $\gamma$ (see
Eq.~(\ref{ICTcNA1}) and above) and from Eq.~(\ref{ICTcNA2}) one
obtains the non-adiabatic contribution to the IC of the penetration
depth in Pr-doped and O-depleted YBCO. One obtains $\beta_{na} =
-0.08$ in the first case and $\beta_{na} = -0.14$ in the second.

Using Eqs.~(\ref{ICTcMI}), (\ref{ICdeltaMI}) and (\ref{ICTcNA1}), as
well as the values of the parameters $\gamma$ and $\tilde{\Gamma}_s$
given above, one can calculate the IC resulting from
the contributions of magnetic impurities and non-adiabaticity for
YPrBCO near $T_c$. The result is shown in Fig.~10 for three different
temperatures.
\unitlength1cm
\begin{figure}[htb]
\begin{center}
\input{fig3AB.ps}
\end{center}
{\small {\bf Figure 10.} Dependence of the isotope coefficient $\tilde{\beta}_{m+na}$ on $T_c$ for $T/T_c = 0.75$ (solid line), $0.85$ (dashed), $0.95$ (dotted). $\alpha_0 = 0.025$, $\gamma = 0.16$, and $\tilde{\Gamma}_s = 123$K (parameters for Y$_{1-x}$Pr$_x$Ba$_2$Cu$_3$O$_{7-\delta}$).}
\end{figure}
One notes that contrary to the case of magnetic
impurities alone (one sets $\gamma = 0$) studied in
Sec.~\ref{sec:MI} (Figs.~2 and 3), the isotope effect of the
penetration depth does
not have the same qualitative behaviour as the IC of $T_c$ (compare
Figs.~9 and 10). The change of curvature does not take place at
the same value of $T_c$ (in the case of $\tilde{\beta}_{m+na}$ the
change occurs at high $T_c$'s and is barely visible on the figure;
see Ref.~\cite{BKW2} for a discussion of this point). One should
measure the IC of $\delta$ for YPrBCO and YBCO, since
it would allow one to give a better estimate of the parameters and
to test the theory.

\section{CONCLUSIONS}\label{sec:concl}

We have reviewed different aspects of the isotope effect. In a first
part, we have summarized the main theories developped to explain
the isotope effect in conventional superconductors and we have
discussed
their relevance for high-$T_c$ oxides. In particular, we have
considered the effect of the Coulomb interaction, band structure
(van Hove singularities),
multiatomic compounds, anharmonicity and non-phononic mechanisms
on the value of the isotope coefficient. These
effects can account for most experimental data obtained on
conventional superconductors. For example, they allow one to
describe the deviation from the standard BCS value $\alpha=0.5$ in
transition metals or the inverse isotope effect ($\alpha<0$) found
in PdH. The results show clearly that a vanishing or even a negative
isotope coefficient can be obtained within conventional
superconductivity where the pairing between charge-carriers is
mediated by the electron-phonon interaction. Generally, however, the
small isotope coefficient is obtained only in conjunction with low
$T_c$'s.
On the other hand, one can also obtain values of the isotope
coefficient that are larger than $\alpha=0.5$ when anharmonic
or band-structure effects are present in the system.

In a second part, we have analyzed the effect of magnetic impurities,
proximity contacts and non-adiabatic charge-transfer processes on
the isotope coefficient (see also
Refs.~\cite{KW1,KBWO,BKW1,BKW2,KW2}). An important common feature
of these factors is that they are not related to the pairing
mechanism but, nevertheless, strongly affect the isotope effect.
On notes that $\alpha>0.5$ is allowed in all three channels.
Furthermore, these factors can induce a non-trivial isotopic shift of
quantities such as the penetration depth $\delta$.
In the case of magnetic impurities and the proximity effect, the
isotope effect of the penetration depth is temperature dependent.
This phenomenon has not been investigated experimentally yet.
In the presence of non-adiabaticity, we have established a
relation between the isotope shift of $T_c$ and $\delta$ for
London superconductors.

The theory presented in the second part of this review allowed us to
describe the unconventional behavior of the isotope coefficient in
YBa$_2$Cu$_3$O$_7$ related systems (Zn and Pr-substituted as well
as oxygen-depleted materials). The case of Zn-substituted YBCO
was described by involving solely the magnetic impurity channel
and did not require any fitting parameter. The theoretical curve
also applies to conventional superconductors.

Our calculations suggest several experiments both on conventional
and high-temperature superconductors. In particular, it would be
interesting to measure the change of the isotope coefficient
induced by a proximity system or magnetic impurities in
conventional superconductors to test our theory. Furthermore,
more experiments have to be done on high-temperature superconductors
so as to determine more precisely the parameters appearing in the
theory.

\section{Acknowledgment}

A.B.~is grateful to the Naval Research Laboratory and the Swiss
National Science Foundation for the support. The work of V.Z.K.~is
supported by the U.S.~Office of Naval Research under contract
No.~N00014-96-F0006.

\section{APPENDIX}

We derive the explicit form of the function $R_0$ given in
Eq.~(\ref{ICdeltaMI0}) (see Ref.~\cite{BKW2}). The
penetration depth
calculated by Skalski {\it et al.} at zero temperature is given by
$\delta^{-2} = -(4\pi n e^2/mc^2)\tilde{K}(\omega=0,{\bf q}=0)$
with\cite{skalski}
\begin{eqnarray}\label{K0sl}
\tilde{K}(0,0) = -\frac{1+\bar{\Gamma}\bar{\eta}^{-3}}{\bar{\eta}}
\left[
\frac{\pi}{2}-\frac{f(\bar{\eta})}{R(\bar{\eta})}
\right]
+
\bar{\Gamma}\bar{\eta}^{-3}\left[
\frac{2}{3}\bar{\eta}-\frac{\pi}{4}\bar{\eta}+1
\right]
\end{eqnarray}
for $\bar{\Gamma},\bar{\eta}<1$ (with
$f(\bar{\eta})=\textrm{arcos}\bar{\eta}$) or
$\bar{\Gamma}<1$, $\bar{\eta}>1$ (with
$f(\bar{\eta})=\textrm{arcosh}\bar{\eta}$) and
\begin{eqnarray}\label{K0ll}
\tilde{K}(0,0) &=& - \frac{1+\bar{\Gamma}\bar{\eta}^{-3}}{\bar{\eta}}
\left\{
\frac{\pi}{2} - 2\frac{\bar{\Gamma}-1}{R(\bar{\Gamma})}
- R^{-1}(\bar{\eta})
  \left( \textrm{arcosh}\bar{\eta} - 2\textrm{artanh}{\cal R} \right)
\right\} \\
&&+ \bar{\eta}^{-3}
\left\{
   \left(\frac{2}{3}\bar{\eta}^2 + 1\right)
   \left(\bar{\Gamma} - R(\bar{\Gamma})\right)
   - \frac{1}{2}\eta R(\bar{\Gamma})\left(\frac{2}{3}\eta - 1\right)
   - \bar{\eta}\bar{\Gamma}\left( \frac{\pi}{4}
       - \frac{\bar{\Gamma}-1}{R(\bar{\Gamma})} \right)
\right\}
\nonumber
\end{eqnarray}
for $\bar{\Gamma},\bar{\eta}>1$. We have introduced the notation
$\bar{\Gamma} = \Gamma_s/\Delta$,
$\bar{\eta} = \eta\bar{\Gamma} = \Gamma_2/\Delta$,
$\eta = \Gamma_2/\Gamma_s$,
$R(x) = \sqrt{|1-x^2|}$ with $x = \bar{\Gamma},\bar{\eta}$ and
${\cal R} = [(\bar{\Gamma}-1)(\bar{\eta}-1)
/(\bar{\Gamma}+1)(\bar{\eta}+1)]^{1/2}$.
$\Delta\equiv\Delta(T=0,\Gamma_s)$ is the order parameter in the
presence of magnetic impurities.
Eqs.~(\ref{K0sl}),(\ref{K0ll}) are valid when $\Gamma_s\ll\Gamma_2$.
These two scattering amplitudes, $\Gamma_2$
and $\Gamma_s$ ($\Gamma_s = \Gamma_1 - \Gamma_2$), defined by
Abrikosov and Gor'kov\cite{AG}, describe the direct and exchange
scattering, respectively.
One can calculate the magnetic impurity contribution to the IC at
$T=0$ from Eq.~(\ref{ICdelta}) in a straightforward way using
Eqs.~(\ref{K0sl}) and (\ref{K0ll}). The result can be written as
\begin{eqnarray}\label{icMI0}
\beta_{m}(T=0) =
-\frac{\alpha_{\Delta}}{2}\frac{K_1+K_2}{\tilde{K}(0,0)}
\end{eqnarray}
where $\alpha_{\Delta}$ is defined below, $\tilde{K}(0,0)$ is
given by Eqs.~(\ref{K0sl}), (\ref{K0ll}) and
\begin{eqnarray}\label{ICsl}
K_1 &=& - \frac{1+3\bar{\Gamma}\bar{\eta}^{-3}}{\bar{\eta}}
       \left(\frac{\pi}{2}-\frac{f(\bar{\eta})}{R(\bar{\eta})}\right)
        \pm \frac{1+\bar{\Gamma}\bar{\eta}^{-3}}{R(\bar{\eta})^2}
     \left(\frac{\bar{\eta}}{R(\bar{\eta})}f(\bar{\eta})-1\right)
\\
K_2 &=& \bar{\Gamma}\bar{\eta}^{-3}
        \left(2 - \frac{\pi}{4}\bar{\eta}\right)
\nonumber
\end{eqnarray}
for $\bar{\Gamma},\bar{\eta}<1$ (upper sign,
$f(\bar{\eta})=\textrm{arcos}\bar{\eta}$) or
$\bar{\Gamma}<1$, $\bar{\eta}>1$ (lower sign,
$f(\bar{\eta})=\textrm{arcosh}\bar{\eta}$) and
\begin{eqnarray}\label{ICll}
K_1 &=& 
- \frac{1+3\bar{\Gamma}\bar{\eta}^{-3}}{\bar{\eta}}
\left\{
\frac{\pi}{2} - 2\frac{\bar{\Gamma}-1}{R(\bar{\Gamma})}
- \frac{1}{R(\bar{\eta})}
\left(\textrm{arcosh}\bar{\eta} - 2\textrm{artanh}\cal{R}\right)
\right\}
- \frac{1+\bar{\Gamma}\bar{\eta}^{-3}}{\bar{\eta}}
\left\{
   2\frac{\bar{\Gamma}(\bar{\Gamma}-1)}{R(\bar{\Gamma})^3}
\right.
     \nonumber\\
&& \left. 
- \frac{1}{R(\bar{\eta})}
\left[
     \frac{\bar{\eta}}{R(\bar{\eta})^2}
     \left(\textrm{arcosh}\bar{\eta} - 2\textrm{artanh}\cal{R}\right)
   + \frac{{\cal R}}{1 - {\cal R}^2}
     \left(  \frac{\bar{\Gamma}^2}{R(\bar{\Gamma})^2}
           + \frac{\bar{\eta}^2}{R(\bar{\eta})^2}
     \right)
   - \frac{\bar{\eta}}{R(\bar{\eta})}
\right]
\right\} \\
K_2 &=& \frac{3}{\bar{\eta}^3}
          \left( \frac{2}{3}\bar{\eta}^2 + 1 \right)
                [\bar{\Gamma}- R(\bar{\Gamma})]
- \frac{1}{2} \eta R(\bar{\Gamma})
  \left( \frac{2}{3}\eta^2 - 1\right)
- \bar{\eta}\bar{\Gamma}
  \left( \frac{\pi}{4} - \frac{\bar{\Gamma} - 1}{R(\bar{\Gamma})}
  \right) \nonumber\\
&& + \frac{1}{\bar{\eta}^3}
\left\{
  \left[
     \frac{\bar{\Gamma}}{R(\bar{\Gamma})}
          \left( \frac{2}{3}\bar{\eta}^2 + 1\right)
          - \frac{4}{3}\bar{\eta}^2
  \right]
      [ \bar{\Gamma} - R(\bar{\Gamma}) ]
\right. \nonumber\\
&&
\left.
  + \frac{1}{2}\eta\frac{\bar{\Gamma}^2}{R(\bar{\Gamma})}
      \left( \frac{2}{3}\eta^2 - 1\right)
  + 2\bar{\eta}\bar{\Gamma}
    \left[
       \frac{\pi}{4} - \frac{\bar{\Gamma}-1}{R(\bar{\Gamma})}
          \left( 1 + \frac{\bar{\Gamma}}{2R(\bar{\Gamma})^2}\right)
    \right]
\right\}
\nonumber
\end{eqnarray}
for $\bar{\Gamma},\bar{\eta}>1$ and $\Delta T_c/T_c(\bar{\Gamma}-1)
\ll 1$. The last condition expresses the fact that the calculation
is not valid in the immediate vicinity of $\bar{\Gamma}=1$.
Eq.~(\ref{icMI0}) contains
$\alpha_{\Delta}$ which is the IC of the order parameter $\Delta$.
In strong-coupling systems, $\alpha_{\Delta}$ has to be calculated
numerically using Eliashberg's equations. Here we calculate the IC
in the framework of the BCS model where $\alpha_{\Delta}$ can be
calculated analytically. Indeed, from the relations
\begin{eqnarray}\label{gapAG}
\ln\left(\frac{\Delta}{\Delta_0}\right) =
\left\{ \begin{array}{l@{\quad : \quad}l}
-\frac{\textstyle \pi}{\textstyle 4}{\textstyle \bar{\Gamma}}
& \bar{\Gamma}\le 1 \\
-\ln\left[\bar{\Gamma} + R(\bar{\Gamma}) \right]
+ \frac{^{\textstyle R(\bar{\Gamma})}}{_{\textstyle 2\bar{\Gamma}}}
- \frac{^{\textstyle \bar{\Gamma}}}{_{\textstyle 2}}
\arctan R(\bar{\Gamma})^{-1}
& \bar{\Gamma} > 1
\end{array}\right.
\qquad ,
\end{eqnarray}
derived by Abrikosov and Gor'kov [$\Delta_0 = \Delta(T=0,\Gamma_s=0)$
is the order parameter in the absence of magnetic impurities] one
obtains
\begin{eqnarray}\label{ICgap}
\alpha_{\Delta} = \alpha_{\Delta_0}
\left\{ \begin{array}{l@{\quad : \quad}l}
\left( 1 - \frac{\textstyle \pi}{\textstyle 4}
{\textstyle \bar{\Gamma}} \right)^{-1}
& \bar{\Gamma}\le 1 \\
\left[
1 - \frac{^{\textstyle \bar{\Gamma}}}{_{\textstyle 2}}
\arctan R(\bar{\Gamma})^{-1} -
\frac{^{\textstyle R(\bar{\Gamma})}}{_{\textstyle 2\bar{\Gamma}}}
\right]^{-1}
& \bar{\Gamma} > 1
\end{array}\right.
\qquad .
\end{eqnarray}
In the BCS approximation one further has
$\alpha_{\Delta_0} = \alpha_0$, where the last quantity was defined
before as the IC of $T_{c0}$, that is, in the absence of magnetic
impurities.

Finally, one obtains Eq.~(\ref{ICdeltaMI0}) with $R_0$ given by:
\begin{eqnarray}
R_0 = -\frac{\alpha_{\Delta}}{2\alpha_0}
\frac{K_1+K_2}{\tilde{K}(0,0)}
\end{eqnarray}
and $\alpha_{\Delta}$ is given by by Eq.~(\ref{ICgap}).


\end{document}